\documentstyle[12pt]{article}
\makeatletter
\newdimen\normalarrayskip              % skip between lines
\newdimen\minarrayskip                 % minimal skip between lines
\normalarrayskip\baselineskip
\minarrayskip\jot
\newif\ifold             \oldtrue            \def\new{\oldfalse}
\def\arraymode{\ifold\relax\else\displaystyle\fi} % mode of array enrties
\def\eqnumphantom{\phantom{(\theequation)}}     % right phantom in eqnarray
\def\@arrayskip{\ifold\baselineskip\z@\lineskip\z@
     \else
     \baselineskip\minarrayskip\lineskip2\minarrayskip\fi}
\def\@arrayclassz{\ifcase \@lastchclass \@acolampacol \or
\@ampacol \or \or \or \@addamp \or
   \@acolampacol \or \@firstampfalse \@acol \fi
\edef\@preamble{\@preamble
  \ifcase \@chnum
     \hfil$\relax\arraymode\@sharp$\hfil
     \or $\relax\arraymode\@sharp$\hfil
     \or \hfil$\relax\arraymode\@sharp$\fi}}
\def\@array[#1]#2{\setbox\@arstrutbox=\hbox{\vrule
     height\arraystretch \ht\strutbox
     depth\arraystretch \dp\strutbox
     width\z@}\@mkpream{#2}\edef\@preamble{\halign \noexpand\@halignto
\bgroup \tabskip\z@ \@arstrut \@preamble \tabskip\z@ \cr}%
\let\@startpbox\@@startpbox \let\@endpbox\@@endpbox
  \if #1t\vtop \else \if#1b\vbox \else \vcenter \fi\fi
  \bgroup \let\par\relax
  \let\@sharp##\let\protect\relax
  \@arrayskip\@preamble}
\def\eqnarray{\stepcounter{equation}%
              \let\@currentlabel=\theequation
              \global\@eqnswtrue
              \global\@eqcnt\z@
              \tabskip\@centering
              \let\\=\@eqncr
              $$%
 \halign to \displaywidth\bgroup
    \eqnumphantom\@eqnrel\hskip\@centering
    $\displaystyle \tabskip\z@ {##}$%
    &\global\@eqcnt\@ne \hskip 2\arraycolsep
         $\displaystyle\arraymode{##}$\hfil
    &\global\@eqcnt\tw@ \hskip 2\arraycolsep
         $\displaystyle\tabskip\z@{##}$\hfil
         \tabskip\@centering
    &{##}\tabskip\z@\cr}
\makeatother

\def\beq{\begin{equation}}
\def\eeq{\end{equation}}
\def\bea{\begin{eqnarray}}
\def\eea{\end{eqnarray}}
\def\stackreb#1#2{\mathrel{\mathop{#2}\limits_{#1}}}

\def\nn{\nonumber}

\begin{document}

\begin{titlepage}
\begin{center}
%{\it P.N.Lebedev Institute preprint}
 \hfill FIAN/TD-19/93

%{\it I.E.Tamm Theory Department}
 \hfill UUITP-7/94
\begin{flushright}{hepth@xxx/???????}\end{flushright}
\begin{flushright}{April 1994}\end{flushright}
\vspace{0.1in}{\Large\bf  String theory and classical integrable
systems}\\[.4in]
{\large A.Marshakov
\footnote{lecture given at III Baltic Rim student seminar, Helsinki,
September 1993}
\footnote{E-mail address:
mars@td.fian.free.net \ \ marshakov@nbivax.nbi.dk \ \
andrei@rhea.teorfys.uu.se}
}\\
\bigskip {\it  Theory Department \\ P.N.Lebedev Physics
Institute \\ Leninsky prospect, 53, Moscow, 117 924, Russia,
\\and\\
Institute of Theoretical Physics, Uppsala University, Uppsala, Sweden}

\end{center}
\bigskip \bigskip

\begin{abstract}
We discuss different formulations and approaches to string theory and  $ 2d$
quantum gravity. The generic idea to get a unique description of {\it many}
different string vacua altogether is demonstrated on the examples in $ 2d$
conformal, topological and matrix formulations. The last one naturally brings
us to the appearance of classical integrable systems in string theory.
Physical meaning of the appearing structures is discussed and some attempts
to find directions of generalizations to ``higher-dimensional" models are made.
We also speculate on the possible appearence of quantum integrable structures
in string theory.

\end{abstract}

\end{titlepage}

\newpage
\setcounter{footnote}0

\section{Introduction}

String theory or theory of $2d$ gravity continues to be one of the main
directions of investigation in mathematical physics. Recent years have brought
us to some progress in understanding its relation to a much older field of
interest
for many mathematical physicists - integrable systems. At the moment we can
already advocate that partially ``integrable science" is directly related to
string theory -- that part connected with {\it classical} integrable equations
and their hierarchies. The situation with {\it quantum} integrable systems
is not yet as clear
\footnote{though there are already many arguments making us beleive that
quantum
integrable systems should play an essential role in formulation of string
theory.}
 so I will almost skip this question below (except for some minor speculations
at the end).
In contrast, the appearance of hierarchies of classical integrable equations
in description of non-perturbative string amplitudes is already a well-known
fact at least for low-dimensional string models
\cite{Douglas,FKN1,DVV1,GMMMO,KMMMZ}.

Below I will try to stress the most essential points of this formalism when the
hierarchies of integrable equations appear in string theory and
discuss the parallels with other languages. I will also
try to clarify the target-space picture for these models and demonstrating
possible  ``generalizations"
to the case of ``higher-dimensional" string theories.

Let me remind you now some questions that can appear in the modern string
theory from a physical point of view
\footnote{with a hope that the word "physics" is right in this context}.
If we beleive that string theory could have something to do with our reality
then the idea is to find a convenient language suitable for computation of
physical quantities - physical amplitudes. It would be marvelous if they could
be computed exactly and in {\it any} background. Unfortunately
nobody knows how to do this for most of string theories with the exception of
those where amazing structure of integrable equations has appeared. However,
there is no complete effective target space theory even for these models - if
exists such a theory will be a good candidate for the role of string field
theory \cite{WittSFT,Zwiebach,M}.

The simplest example of string theory (mostly well-known) is topological pure
gravity (which has lots of different equivalent formulations and is mostly
well-studied)\cite{Witten,Distler,GMM,K1}. Naively such theory should not have
target space at all. Then the natural question is what is the origin of the
nice structure appearing in the form of the Virasoro constraints acting to a
partition function \cite{WittK,MMM,Mikh}, the set of which is actually
equivalent to the concept of integrability in these models?

Below we are going to pay attention to several points concerning integrability
arising in description of low-dimensional string theory and try to show how the
structure of hierarchies of classical integrable equations can be generalized
to higher dimensional theories. The main idea is that the phase space of
appearing classical integrable models may be considered as a phase space for
effective string field theory, and its quantization can lead to the formulation
of the second-quantized string theory in terms of (quantum?) integrable
systems.

First, we review a little the $ 2d$ conformal field theory language -- the
basic Polyakov definition of string theory. We will concentrate on Liouville
(physical) gravity coupled to most known non-critical string -- so called
$(p,q)$ models, and try to discuss its target-space structure when taking the
quasiclassical limit. As in any covariant description this one requires a lot
of ``extra" information of the theory, this leads to the situation when the
original formulation of string theory is not effective for answering questions
about its target-space structure etc (like it occurs in the theory of
particles). We are going to consider an example of motion in the space of $
(p,q)$ theories - a step towards their effective description and then turn to
the other ways to formulate the same theories.

At least part of these models ($ (p,1)$ and/or $ (1,p)$) allow to consider them
by topological theory language \cite{DVVLG,Dijkgraaf,LP}
\footnote{see also A.Losev's contribution to this volume}
. Coupling to reparameterization ghosts one gets total $c=0$ central charge
which allows an interpretation
in terms of the twisted $N=2$ theory \cite{Li}. Physical gravity is now
included in
``topological matter" -- while topological gravity appears roughly speaking in
the integration over module space - thus, generalization of the notion of {\it
critical} string. In principle even
$26$-dimensional bosonic string can be considered as topological theory, but
low-dimensional examples allow one to demonstrate better the explicit $N=2$
cancellation
of bosonic and fermionic two-dimensional determinants and target-space
co-ordinates appear as corresponding zero modes. The result is very close to
{\it localization} formulas appeared to be a useful tool when studying
topological, quantum-mechanical and integrable models \cite{loc}.

It appears that at least for topological models there exists a very effective
exact formulation on the language of matrix models. Indeed, the matrix integral
is a sort of counting of possible two-dimensional diagrams thus being a natural
object in the theory of pure gravity - or string theory with an empty target
space. These integrals naturally come from an intersection theory on module
space \cite{K1}, what would be more interesting is if one can successfully
realize the same idea for the moduli of target-spaces (see recent papers
\cite{K2,KoMa}).

Matrix models brought us to the understanding of possible role of integrable
systems in string theory. Till the moment only for these ``empty models" but
there exists a way to compute the exact amplitudes in these string theories by
prooving that their generation function is a $ \tau$-function of KP (or in
general Toda-lattice) hierarchy satisfying some natural additional constraint.
This constraint is an analog of unitarity condition and can be also interpreted
in terms of some flow in the space of different low-dimensional models.

However, there are some puzzles, arising along this way. They are directly
connected with the question of interpretation of  $p-q$ duality, which is in
order repated to the problem of string field theory background independence
\cite{WittBI}. We are going to discuss them below (see also \cite{KM3}).

\section{2d conformal theories}

Let us start with reminding that by canonical string theory one usually has in
mind the induced two dimensional gravity, having the following form in the
Polyakov's path integral approach:

\begin{equation}\label{Polyakov}
\int{DgD\phi e^{- S[\phi]}} \sim \int{Dg e^{\gamma\int{R{1 \over \triangle}R +
 {\mu}^{2}\sqrt{g} }}}
\end{equation}
where $ \phi$ stands for the integration over some $ 2d$ conformal field theory
in the background world-sheet metric $ g$.

It seems to be true that at the moment there is not still an existing
consistent method of quantization the appearing in (\ref{Polyakov}) Liouville
theory (see however \cite{Gervais,GerSch} etc). The common beleif is that the
following ideology is right \cite{KPZ,DK,D}. Consider the integral in
(\ref{Polyakov}) as integral over {\it conformal field theory} consisting of $
(i)$ conformal matter; $ (ii)$ reparameterization ghosts $ b$ and $ c$;
$ (iii)$ {\it conformal} Liouville theory. The latter one should be {\it
defined} as a conformal theory with the central charge $ 26 - c_{matter}$ in
order to cancell the anomaly.

For the simplest example of so-called $(p,q)$ models
coupled to $2d$ gravity $ \phi$ can be simply considered as a "deformed"
scalar field. Choosing the
gauge for metric $ g_{ab} = e^{\varphi}\hat{g}_{ab}$ we reduce the problem to a
conformal theory of two fields $ \phi$ and $ \varphi$ with the stress-tensors

\begin{equation}
T_m = - {1 \over 2}\left( \partial\phi \right)^2 +
i\alpha_0\partial ^2 \phi
$$
$$
T_L =  - {1 \over 2}\left( \partial\varphi \right)^2 +
\beta_0\partial ^2 \varphi
\end{equation}
where

\begin{equation}
\partial\phi (z) \partial\phi (0) = - {1 \over z^2} + ...
$$
$$
\partial\varphi (z) \partial\varphi (0) = - {1 \over z^2} + ...
\end{equation}
thus giving the Virasoro central charges

\begin{equation}
c_m = 1 - 12\alpha _0^2
$$
$$
c_L = 1 + 12\beta _0^2
\end{equation}
satisfying $ c_m + c_L - 26 = 0$ with $ -26$ coming from the reparameterization
ghosts contribution. For the $ (p,q)$ theories

\begin{equation}
\alpha_0 = \sqrt{{p \over 2q}} - \sqrt{{q \over 2p}}
$$
$$
\beta_0 =   \sqrt{{p \over 2q}} + \sqrt{{q \over 2p}}
\end{equation}
or

\begin{equation}
\beta_0 = \sqrt{2}\cosh{\theta}
$$
$$
\alpha_0 = \sqrt{2} \sinh{\theta}
\end{equation}
with

\begin{equation}
\theta = {1 \over 2}\log{{p \over q}}
\end{equation}
For such system one has the following matter Kac spectrum

\begin{equation}\label{alpha}
\alpha_+ = \sqrt{{2p \over q}}
$$
$$
\alpha_- = - \sqrt{{2q \over p}}
$$
$$
\alpha_{n,m}  = {1-n \over 2}\alpha_+
+ {1-m \over 2}\alpha_- = {(1-n)p - (1-m)q \over \sqrt{2pq} }
$$
$$
\Delta_{n,m} = {(np-mq)^2 - (p-q)^2 \over 4pq}
$$
$$
\Delta_{min} = {1 - (p-q)^2 \over 4pq}
\end{equation}
where in matter sector there exists a ``periodicity" allowing one to restrict
to
\footnote{and actually this ``minimality" is broken by interacting with gravity
(see for example \cite{Dots})}

\begin{equation}
n = 1,...,q-1
$$
$$
m = 1,...,p-1
\end{equation}
while the gravity sector is given by

\begin{equation}\label{beta}
\beta_{\pm} = \pm\alpha_{\pm}
$$
$$
\beta_{n,m} ={p+q \pm (np-mq) \over \sqrt{2pq} } \rightarrow {(1-n)p + (1+m)q
\over \sqrt{2pq} }
$$
$$
\beta _{min} = {p+q \pm 1 \over \sqrt{2pq} }
\end{equation}
where we have chosen a sign in order to make correspondence to the proper
quasiclassical limit.

Indeed, we see that the conformal $ (p,q)$ model is totally symmetric under
exchange of $ p$ and $ q$. However, the difference between $ p$ and $ q$
becomes crucial when coupling to $2d$ gravity, or better to say when
considering string theory. In fact, the asymetry appears if one takes the
quasiclassical limit
\footnote{this quasiclassical limit is important if we want to discuss
target-space properties of the theory}
then only half of the screening operators have well-defined limit ($ \alpha
_{+}$ and $ \beta _{+}$ for $ q \rightarrow \infty$ and $ \alpha _{-}$ and $
\beta _{-}$ for $ p \rightarrow \infty$). The simplest way to see it is to
consider $ (p,q)$ model as a Hamiltonian reduction of the WZNW theory
\cite{GMMhr} and for the WZNW theory it is known \cite{GMMOS} that only one
screening operator (having smooth limit for $ k \rightarrow\infty$) appears
naturally from classical action as a constraint on free fields. Physically it
means that one has to choose the operator coupling to unity (or lowest
dimensional one) - i.e. what is called the puncture operator in a theory.

\subsection{Rotations in the space of free fields: the way to move in the space
of theories}

Now, let us turn to the question how one can describe the whole set of
different $ (p,q)$ models. In fact this is rather hard to do using the
technique of present section - more or less complete descriprion exists only
based on the methods presented below. However, here we will try to use as much
as possible of conformal methods in order to get understanding of possible
flows in the space of theories.

One can consider the following rotation in the space of $ (p,q)$ theories

\begin{equation}
\tilde{\beta_0} = \alpha_0 \sinh{\vartheta} + \beta_0 \cosh{\vartheta}
$$
$$
\tilde{\alpha_0} = \alpha_0 \cosh{\vartheta} + \beta_0 \sinh{\vartheta}
\end{equation}
The same rule one has for primary operators $e^{i\alpha\phi + \beta\varphi}$,
labeled by (\ref{alpha}), (\ref{beta})

\begin{equation}
\tilde{\beta} = \alpha \sinh{\vartheta} + \beta \cosh{\vartheta}
$$
$$
\tilde{\alpha} = \alpha \cosh{\vartheta} + \beta\sinh{\vartheta}
\end{equation}
with parameter of the transformation

\begin{equation}
\vartheta = {1 \over 2}\log{{\tilde{p} \over \tilde{q}} {q \over p}}
\end{equation}
Now it is easy to rewrite it in the space of fields
\footnote{Note that in particular such rotation makes from {\it real} Liouville
field for $ c=1$ {\it complex-valued} for $ c<1$.}

\begin{equation}
\Phi _{n,m} = \exp{\left(i\alpha _{n,m}\phi + \beta _{n,m}\varphi\right)}
\end{equation}
One finds that for $ \Phi _{n,m}^{(p,q)} \rightarrow  \Phi _{\tilde n,\tilde
m}^{(\tilde p,\tilde q)} $

\begin{equation}\label{nmmove}
\tilde{n} = {p \over q}n
$$
$$
\tilde{m} = m
\end{equation}

Now, let us consider an explicit example how the transformation (\ref{nmmove})
works. First, let us take (\ref{beta}) and make there a substitution

\begin{equation}\label{krqp}
kq + r = np-mq
\end{equation}
for $(q,p)$ theory and

\begin{equation}\label{krpq}
kp + r  = np-mq
\end{equation}
for $(p,q)$ theory. Then (we restrict ourselves to the second choice
(\ref{krpq}))

\begin{equation}
k = n - \left[ {qm \over p}\right]
$$
$$
r = mq - p\left[{qm \over p}\right]
\end{equation}
where $[x]$ means integer part of $x$, and this give the correspondence
\cite{MSS}

\begin{equation}\label{MSS}
\sigma _k ({\sl O}_r) \sim \int{\exp{\left(i\alpha _{n,m}\phi + \beta
_{n,m}\varphi\right)}}
\end{equation}
i.e. $ r$ enumerates ``topological primary fields" and we have defined
(\ref{krqp}) and (\ref{krpq}) in order to have exactly $ q-1$ or $ p-1$ of
them. In such terminology $ k$ counts their ``gravitational descendants".

Now the transformation (\ref{nmmove}) works as follows: take for example $
(p,1)$ theory and consider $ \sigma _1 ({\bf1})$. $ \bf 1$ is given by
zero-dimensional matter operator with

\begin{equation}
r = m = p-1
$$
$$
k = n = 0
$$
$$
\Delta_{\bf 1} = \Delta _{0,p-1} = 0
\end{equation}
while for $ \sigma _1 ({\bf1})$ itself one has

\begin{equation}
r = m = p-1
$$
$$
k = n = 1
\end{equation}
Then, making transformation (\ref{nmmove}) we get

\begin{equation}
\tilde{m} = m = p-1
$$
$$
\tilde{n} = p
\end{equation}
It means that the rotated field becomes {\it primary} one in the $
(\tilde{p},\tilde{q})$ model with $\tilde{p} = p$ and

\begin{equation}
\tilde{k} = \tilde{n} -  \left[ {\tilde{q}m \over p}\right] = 0
\end{equation}
which gives

\begin{equation}
\tilde{q} = p + \left[{\tilde{q} \over p}\right]
\end{equation}
or just $ \tilde{q} = p+1$. For $p=2$ such an operator drops us from pure
topological gravity $(p,q) = (2,1)$ to the pure physical gravity point
$(\tilde{p},\tilde{q}) = (2,3)$.

The example considered above as just an illustration of the flow in the space
of simplest string theories. We have seen that in the original conformal
formulation they strongly depend on the basis one has to choose in the space of
fields and/or observables. That is one of the reasons why more convenient
target space description for string theory is necessary. On the language of
matrix models these relations can be rewritten in the form of the Virasoro-$ W$
constraints and generalized KdV flows. We will see below, that the effective
target-space description based on integrable systems gives much stronger
possibilities to investigate this phenomenon.

\section{Topological language}

Now, let us make a sort of an intermideate step -- to reformulate the above
picture in the following way. Forget about the difference between conformal
matter and conformal gravity since metric contrubution -- Liouville and ghost
sectors are also represented by certain conformal theories. Then one can
generalize the above consideration restricting to the only requirement that the
total central charge of a theory is equal to zero. The presence of gravity
remains in the only fact that the result after all should be integrated over
{\it module} space. Such object is usually meant by what is called {\it
topological gravity} \cite{Witten,LP}. From such point of view critical string
is a good example of a topological theory interacting with topological gravity
except for the only case that integral over module space might diverge.

Now consider this (conformal matter plus Liouville
gravity plus reparameterization ghosts)

\begin{equation}
T_{gh} = - 2b\partial c + c\partial b
\end{equation}
($T = -jb\partial c - (1-j)c\partial b$ for $j=2$)
as a twisted $ N=2$ superconformal theory \cite{Li}.
In such case for $ (q,p)$ and $ (p,q)$ ``untwisted" models two values of the
central charge are

\begin{equation}\label{cencha}
c _{(q,p)}= 3(1 - {2p \over q})
$$
$$
c_{(p,q)} = 3(1 - {2q \over p})
\end{equation}
This can be demonstrated for example as follows. First let us consider the $
SU(2)_k$ WZNW model. Such theory posseses conformal symmetry with the Virasoro
central charge

\beq\label{su2}
c_{SU(2)_k} = {3k\over k+2}
\eeq
One of the possible ways to get matter $ (p,q)$ model is via the
Drinfeld-Sokolov reduction. Then, one easily finds that

\beq\label{level}
k_{(q,p)} \equiv \tilde{k} = {q\over p} - 2
$$
$$
k_{(p,q)}\equiv k = {p\over q} - 2
\eeq
where asymmetry between $ p$ and $ q$ appeared exactly as we mentioned above
when one has to distinguish the {\it classical} screening operator. The
relation

\beq
k+2 = {1\over \tilde{k} + 2}
\eeq
in particular demonstrates the duality between two ``classical" limits
when $k \rightarrow \infty$ corresponds to $\tilde k \rightarrow -2$ and
vice versa. It can also clarify what is the meaning of the Wess-Zumino model
with a {\it rational} central charge -- considering it as a dual to that one
with integer $ k$ in the above sense.

The most easy way to check the relations (\ref{cencha}), (\ref{level}) is using
bosonization technique when performing the reduction \cite{GMMhr}. Indeed,
``twisting"

\begin{equation}
T_{WZNW} \rightarrow \tilde{T}_{WZNW} = T_{WZNW} - \partial H
\end{equation}
where (see \cite{GMMOS} for more detailed description of free field technique)

\begin{equation}
T_{WZNW} = w\partial\chi - {1 \over 2}(\partial\phi)^2 - {i \over \sqrt{2(k+2)}
}\partial ^2 \phi
$$
$$
H = w\chi - {i \over \sqrt{2} }\sqrt{k+2}\partial\phi
\end{equation}
one gets

\begin{equation}\label{WZtw}
\tilde{T}_{WZNW} = -\partial w \chi - {1 \over 2}(\partial\phi)^2 - {i \over
\sqrt{2} }\left({1 \over \sqrt{k+2} } - \sqrt{k+2}\right)\partial ^2\phi
\end{equation}
and the field $ \phi$ stands now for minimal model, i.e. the corresponding
screening operators become the screening charges of the $ (p,q)$ model.

Another valuable relation exists between the $ SU(2)_k$ WZNW model and $ N=2$
minimal model $ A_k$. Namely the $ SU(2)_k$ Kac-Moody currents can be
represented as

\begin{equation}
J_{\pm} = e^{\pm i \sqrt{{2 \over k}}h \mp i \sqrt{1+{2 \over k}}\Phi  }G_{\pm}
\end{equation}
with

\begin{equation}
H = i\sqrt{{k \over 2}}\partial h
\end{equation}
-- the Cartan current of the $ SU(2)_k$ while

\begin{equation}
J = i\sqrt{{k \over k+2}}\partial\Phi
\end{equation}
being the $ U(1)$ current of the $ N=2$ minimal model and $ G_{\pm}$ denote the
corresponding superconformal symmmetry generators. The twisting of $ N=2$ gives

\begin{equation}\label{N2tw}
T_{N=2} \rightarrow T_{N=2}^{tw} = T_{N=2} - {i \over 2}\sqrt{{k \over
k+2}}\partial ^2 \Phi
\end{equation}
where
\footnote{the equality should be understood schematically, i.e. in the sense of
bosonization.}

\begin{equation}\label{N2WZ}
T_{N=2} = T_{WZNW} - T_h + T_{\Phi}
\end{equation}
Eqs. (\ref{WZtw}), (\ref{N2tw}) and (\ref{N2WZ}) altogether give

\begin{equation}
T_{N=2}^{tw} = T_{WZNW}^{tw} + \hat{T}_{\Phi} - \hat{T}_h
\end{equation}
where from the first term in the r.h.s. one can single out the $ (p,q)$ matter
model, while the rest can be transformed by similiar technique into the
Liuoville and ghost contributions.

\subsection{Landau-Ginzburg models}

The particular class of topological theories which includes $N=2$
superconformal minimal models is given by the Landau-Ginzburg models. The
action can be written in the form:

\begin{equation}\label{LanGinz}
\int{\partial X \bar \partial X^{\ast} + \psi \bar \partial \psi^{\ast} +
\bar{\psi}\partial\bar{\psi}^{\ast} + FF^{\ast} + W'(X)F +
\psi\bar{\psi}W''(X) + W'(X^{\ast})F^{\ast} +
\psi^{\ast}\bar{\psi}^{\ast}W''(X^{\ast})}
\end{equation}
which is invariant under the $ N=2$ supersymmetry transformations, generated by

\begin{equation}
G = \psi{\delta \over \delta X} + F{\delta \over \delta\bar\psi} - \partial
X^{\ast}{\delta \over \delta \psi^{\ast}} - \partial\bar{\psi}^{\ast} {\delta
\over \delta F^{\ast}}
$$
$$
\bar G = \bar{\psi}{\delta \over \delta X} - F{\delta \over \delta\psi} -
\bar\partial X^{\ast}{\delta \over \delta \bar\psi^{\ast}} +
\bar\partial{\psi}^{\ast} {\delta \over \delta F^{\ast}}
$$
$$
G^{\ast} = \psi^{\ast}{\delta \over \delta X^{\ast}} + F^{\ast}{\delta \over
\delta\bar\psi^{\ast}} - \partial X{\delta \over \delta \psi} -
\partial\bar{\psi} {\delta \over \delta F}
$$
$$
\bar G^{\ast} = \bar{\psi}^{\ast}{\delta \over \delta X^{\ast}} -
F^{\ast}{\delta \over \delta\psi^{\ast}} - \bar\partial X{\delta \over \delta
\bar\psi} + \bar\partial{\psi} {\delta \over \delta F}
$$
$$
\{G,G^{\ast}\} = -2\partial
$$
$$
\{\bar{G},\bar{G}^{\ast}\} = -2\bar{\partial}
\end{equation}
After twisting, the lagrangian takes the form

\begin{equation}\label{LanGintw}
\int{\partial X \bar \partial X^{\ast} + \psi \bar \partial \psi^{\ast} +
\bar{\psi}\partial\bar{\psi}^{\ast} + FF^{\ast} + \psi\bar{\psi}W''(X) + FW'(X)
+ \sqrt{g}\left[F^{\ast}W'(X^{\ast}) +
\psi^{\ast}\bar{\psi}^{\ast}W''(X^{\ast})\right] } =
$$
$$
= \int{{1 \over 2}\psi\bar{\partial}(\psi^{\ast} - \bar{\psi}^{\ast}) + {1
\over 2}\bar{\psi}\partial (\bar{\psi}^{\ast} - \psi ^{\ast} )+
\psi\bar{\psi}W''(X) + FW'(X) + \{Q,V\}}
\end{equation}
with

\begin{equation}
Q = G^{\ast} + \bar{G}^{\ast} = \theta{\delta \over \delta X^{\ast}} -
F^{\ast}{\delta \over \delta\eta} - \partial X {\delta \over \delta\psi} -
\bar{\partial}X{\delta \over \delta\bar{\psi}} - (\partial\bar{\psi} +
\bar{\partial}\psi){\delta \over \delta F}
\end{equation}
$ \psi dz$, $ \bar{\psi}d\bar{z}$ and $ Fdzd\bar{z}$ are forms and $ \psi
^{\ast}$, $ \bar{\psi}^{\ast}$ and $ F^{\ast}$ are scalar functions and where

\begin{equation}
V = - \int{ {1 \over 2} (\psi\bar{\partial}X^{\ast} + \bar{\psi} \partial
X^{\ast}) + \sqrt{g}\eta W'(X^{\ast})  }
\end{equation}
where we have introduced

\begin{equation}
\theta = \psi ^{\ast} + \bar{\psi}^{\ast}
$$
$$
\eta = {1 \over 2}(\psi ^{\ast} - \bar{\psi}^{\ast})
\end{equation}
The integral with the action (\ref{LanGintw}) can be computed by localization
technique \cite{loc}. It localizes on $ Q = 0$, i.e.

\begin{equation}\label{loc}
\theta = 0
$$
$$
F^{\ast} = 0 \left( = {\partial W \over \partial X} \right)
$$
$$
\partial X = 0
$$
$$
\bar{\partial}X = 0
$$
$$
\partial\bar{\psi} + \bar{\partial}\psi = 0
\end{equation}

Computation of the path integral for (\ref{LanGintw}) gives zero for the
trivial potential $ W(X) = X$. This is the statement we will use below for $
(1,p)$ models -- stricktly speaking the case $(1,p)$ should correspond to
(\ref{LanGintw}) with a trivial potential and {\it non-trivial} kinetic term,
but the corresponding integral do not depend on kinetic (or $ D$-) term due to
$ N=2$ bosonic-fermionic cancellation. Eqs. (\ref{loc}) demonstrate that
actually the path integral is not still the most effective description for
those models -- it can be reduced to a more simple object. Such objects are
directly related to integrable systems and we will pass to their description
below.

\section{Matrix models}

To understand better the effective description of $ 2d$ gravity and string
models let us for a moment trivialize the situation and return back from
strings to particles, i.e. from surfaces to lines.
A natural question is what is the analog of topological string models in the
one-dimensional case and the answer should be very simple. Indeed, for the
topological one-dimensional theory the only thing which can appear is something
related to the points at the end of paths and their permutations, so these
should be combinatorial numbers attached to the ends of Feynman diagrams.

The module space for one-dimensional theories consists of the lengths of
world-lines, so inclusion of topological one-dimensional gravity should somehow
take this into account. For the simplest case of the propagator one should get

\begin{equation}\label{prop}
G_{\alpha\beta} = \int_{0}^{\infty}{dT f_{\alpha\beta}(T)}
\end{equation}
with $ T$ being the length of the world-line while $ \alpha$ and $ \beta$ are
indices running over the space attached to each point - end of the line, i.e.
over the Hilbert space of the corresponding theory.  The objects $
G_{\alpha\beta} $ can be considered as building blocks for the theory.

In the case of absence of the target space the only choice for $
f_{\alpha\beta}(T)$ is $ \delta_{\alpha\beta}f(T)$, so instead of nontrivial
propagators one gets just a number $ G$ and the "theory" reduces to a
"generation function" via the one-dimensional integral

\begin{equation}\label{1dim}
\int{d\phi \exp{\left(-{{\phi}^{2} \over 2G^2} + t\phi +
\sum{g_n\phi^n}\right)}}
\end{equation}
where one should fix by hands what sort of one-dimensional "branches" - i.e.
geometries is allowed. This is a typical "counting diagram" integral and it
should be considered as a one-dimensional analog of generating function below.

{}From this point of view, two-dimensional topological gravity should naturally
bring to ``fat graphs" where generation function has a nice prescription to be
computed via matrix models. A simple analog of (\ref{1dim}) would look like

\beq\label{mamo}
Z_N = \int DM_{N \times N} \exp \left(- TrV(M)\right)
\eeq
which was proven (in the limit $ N \rightarrow\infty$) to be an effective way
to compute the integral over two-dimensional metrics, including the sume over
topologies. The continuum integration (\ref{Polyakov}) is approximated by
triangulations of world-sheet in (\ref{mamo}).

Below, we will concentrate mostly to a slightly different version of the
integral (\ref{mamo}) which rather has an interpretation of the target space
theory. The exact expression is \cite{KMMMZ}

\beq\label{gkm}
Z^{(N)}[V|M] \equiv  C^{(N)}[V|M] e^{TrV(M)-TrMV'(M)}\int
DX\ e^{-TrV(X)+TrV'(M)X}
\eeq
where the integral is taken over  $N\times N$  ``Hermitean" matrices,
with the normalizing factor given by Gaussian integral

\beq\label{norm}
C^{(N)}[V|M]^{-1} \equiv  \int \hbox{  DY }\ e^{-TrU_2[M,Y]},
$$
$$
U_2 \equiv \lim _{\epsilon \rightarrow 0}
{1\over \epsilon ^2}Tr[V(M+\epsilon Y) - V(M) - \epsilon YV'(M)]
\eeq
Including an external matrix field, which can be considered as a source ($
\equiv$ coupling constants) it allows us do assign more or less concrete
potential to a theory.

The formula (\ref{gkm}) has in fact a lot of similiarities with the
Landau-Ginzburg model we discussed in the previous section. Both theories are
determined by a potential and as we will see below there exists a simple
relation between the potential in (\ref{gkm}) and the superpotential of the
Landau-Ginzburg model $ W(X)$, namely:

\begin{equation}
W(X) = V'(X)
\end{equation}

\subsection{From matrix models to integrable systems}

Now we are going to demonstrate that matrix models being an adequate
formulation for certain very simple string theories
 naturally lead to appearance of the {\it classical} integrable systems
describing the exact solutions for such strings. Namely, we will show that
introduced in the previous section model (\ref{gkm}) is a particular solution
to the KP (Toda lattice) hierarchy. That is:

($A$) The partition function $Z^V_N[M]$ (\ref{gkm}),
if considered as a function of time-variables

\beq\label{miwa}
T_k = {1\over k} Tr\ M^{-k}\hbox{, }  k\geq 1
\eeq
is a KP $\tau$-function for {\it any} value of $N$ and {\it any}
potential $ V [X]$.

($B$) As soon as  ${ V}[X]$  is homogeneous polynomial of degree
$p+1$,
$Z^ {\{V\}}_N[M] = Z^{\{p\}}_N[M]$  is in fact a $\tau $-function of
$p$-reduced KP hierarchy.
\footnote{Moreover, actually,
$\displaystyle {{\partial Z^{\{p\}}\over \partial T_{np}}}=0$.}

In order to prove these statements, first,  we rewrite (\ref{gkm})
in terms of determinant formula

\beq\label{det}
Z^ {\{V\}}_N[M] = {{\rm det} _{(ij)}\phi _i(\mu _j)\over \Delta
(\mu )}\
\ \ \ \ i,j =
1,...,N.
\eeq
Then, we show that {\it any} KP $\tau $-function in the Miwa
parameterization does have the same determinant form.

The main thing which distinguishes matrix models from the point of view of
solutions to the KP-hierarchy is that the set of functions  $\{\phi _i(\mu )\}$
in (\ref{det}) is not arbitrary. This is
the origin of  ${\cal L}_{-1}$ and other  ${\cal W}$- constraints (which in the
context of KP-hierarchy may be considered as implications of  ${\cal L}_{-1})$.

The fact that the classical integrable system appear in string theory, of
course has more deep reason that this simple illustration for low-dimensional
models.

\subsection{Integrability from the determinant formula}

\bigskip
We begin with an evaluation of the integral \cite{KMMMZ}:

\beq
{\cal F}^{\{V\}}_N[\Lambda ] \equiv  \int   DX\ e^{- Tr[ V(X) -
Tr\Lambda X]}
\eeq
The integral over the ''angle" $U(N)$-matrices can be easily taken with the
help of
\cite{IZM}
and if eigenvalues of  $X$  and  $\Lambda $  are denoted by  $\{x_i\}$  and
$\{\lambda _i\}$  respectively, the result is

\beq\label{int}
{1\over \Delta (\Lambda )}\left[  \prod _{i=1}^N
\int   dx_ie^{- V(x_i)+\lambda _ix_i} \right] \Delta (X)
\eeq
$\Delta (X)$  and  $\Delta (\Lambda )$  are Van-der-Monde
determinants, $e.g$. $\Delta (X) = \prod_{i>j}(x_i-x_j)$.

The $r.h.s$. of (\ref{int}) can be rewritten as

\beq
\Delta ^{-1}(\Lambda) \Delta ({\partial \over \partial \Lambda})
\prod _i \int dx_i e^ {- V(x_i) + \lambda _i x_i } =
$$
$$
=
\Delta ^{-1}(\Lambda ) \hbox{det} _{(ij)}F_i(\lambda _j)
\eeq
with

\beq
F_{i+1}(\lambda ) \equiv  \int   dx\ x^ie^{- V(x)+\lambda x} =
({\partial \over \partial \lambda })^iF_1(\lambda ).
\eeq
Note that

\beq
F_1(\lambda ) = {\cal F}^{\{{V}\}}_{N=1}[\lambda ]\hbox{ . }
\eeq
If we recall that

\beq
\Lambda  = V'(M) = W(M)
\eeq
and denote the eigenvalues of  $M$  through  $\{\mu _i\}$ , then:

\beq
{\cal F}^{\{V\}}_N[ W(M)] = {{\rm det} \ \tilde
\Phi _i(\mu _j)\over
\prod _{i>j}(W(\mu _i)- W(\mu _j))}\ ,
\eeq
with

\beq\label{entry1}
\tilde \Phi _i(\mu ) = F_i(W(\mu )).
\eeq
Proceed now to the normalization (\ref{norm}). Indeed, it is given by
the Gaussian integral:

\beq\label{norm1}
C^{(N)}[V|M]^{-1} \equiv  \int   DX\ e^{-U_2(M,X)}.
\eeq
Then for evaluation of
(\ref{norm1}) it remains to use the obvious rule of Gaussian integration,

\beq
\int   DX\ e^{-\sum ^N_{i,j} U_{ij}X_{ij}X_{ji}} \sim \prod ^N_{i,j}
U^{-1/2}_{ij}
\eeq
and substitute the explicit
expression for $U_{ij}(M)$. If potential is represented as a formal series,

\beq
V(X) =\sum  {v_n\over n}X^n
$$
$$
W(X) = \sum{v_n X^n}
\eeq
we have

$$
U_2(M,X)
=\sum ^\infty _{n=0}v_{n+1}\left\lbrace \sum _{a+b=n-1}TrM^aXM^bX
\right\rbrace ,
$$
and

$$
U_{ij} =\sum ^\infty _{n=0}v_{n+1} \left\lbrace
\sum _{a+b=n-1}\mu ^a_i\mu ^b_j \right\rbrace  =
\sum ^\infty _{n=0}v_{n+1} {\mu ^n_i - \mu ^n_j\over \mu _i -
\mu _j}  = {W(\mu _i) -  W(\mu _j)\over \mu _i - \mu _j}.
$$
Coming back to (\ref{gkm}), we conclude that

\beq\label{**}
Z^{\{V\}}_N[M] = e^{Tr[V(M)-MW(M)]}
C^{(N)}[V|M] {\cal F}_N[W(M)] \sim
$$
$$
\sim [{\rm det} \ \tilde \Phi _i(\mu _j)] \prod _{i>j}^N
{U_{ij}\over (W(\mu _i)-W(\mu _j))} \prod _{i=1}
s(\mu _i)  =  {[{\rm det} \ \tilde \Phi _i(\mu _j)]\over \Delta (M)}
\prod _{i=1}^N
s(\mu _i)\ .
\eeq

\beq\label{entry2}
s(\mu ) = [ W'(\mu )]^{1/2} e^{V(\mu )-\mu  W(\mu )}
\eeq
The product of $s$-factors at the $r.h.s$. of (\ref{**}) can be absorbed into
$\tilde \Phi $-functions:

\beq\label{zvdet}
Z^{\{ V\}}_N[M] = {{\rm det}
\Phi _i(\mu _j)\over \Delta (M)}\hbox{,}
\eeq
where

\beq\label{entry3}
\Phi _i(\mu ) = s(\mu )\tilde \Phi _i(\mu )
\stackreb{\mu \to
\infty}{\to} \mu ^{i-1}(1 + {\cal O}({1\over \mu})).
\eeq
where the asymptotic is crucial for the determinant (\ref{zvdet}) to be
a solution to the KP hierarchy in the sense of \cite{SW85}.

\bigskip
{\it The Kac-Schwarz operator} \cite{KS,S}.
{}From eqs.(\ref{entry1}),(\ref{entry2}) and (\ref{entry3}) one can deduce that
$\Phi _i(\mu )$
can be derived from the basic function  $\Phi _1(\mu )$  by the relation

\beq
\Phi _i(\mu ) = [W'(\mu )]^{1/2}\int   x^{i-1}
e^{- V (x) + x  V ' (\mu )}dx =
A_{\{ V\}}^{i-1}(\mu )\Phi _1(\mu )\ ,
\eeq
where  $A_{\{ V\}}(\mu )$  is
the first-order differential operator

\beq\label{ks}
A_{\{ V\}}(\mu )  =  s {\partial \over \partial \lambda } s^{-1} =
{e^{ V(\mu )-\mu  W(m)}\over [W'(\mu )]^{1/2}}
{\partial \over \partial \mu }
{e^{- V(\mu )+\mu  W(\mu )}\over [ W'(\mu )]^{1/2}} =
$$
$$
 =  {1\over  W'(\mu )} {\partial \over \partial \mu } + \mu  -
{ W''(\mu )\over 2[ W'(\mu )]^2}\ .
\eeq
In the particular case of $V(x) = {{x^{p+1}} \over {p+1}}$

\beq
A_{\{p\}}(\mu ) =
{1\over p\mu ^{p-1}} {\partial \over \partial \mu } + \mu  -
{p-1\over 2p\mu ^p}
\eeq
coincides (up to the scale transformation of  $\mu $  and
$A_{\{p\}}(\mu ) )$ with the operator which determines the finite dimensional
subspace
of the Grassmannian in ref.\cite{KS}
We emphasize that the property

\beq
\Phi _{i+1}(\mu ) = A_{\{V\}}(\mu ) \Phi _i(\mu )\ \ \ \
( F _{i+1}(\lambda )
= {\partial \over \partial \lambda } F _i(\lambda ) )
\eeq
is exactly the thing which distinguishes partition functions of GKM from
the expression for generic $\tau $-function in Miwa's coordinates,

\beq\label{taumiwa}
\tau ^{\{ \phi _i\}}_N[M] = {[{\rm det} \ \phi _i(\mu _j)]\over \Delta (M)}
\hbox{,}
\eeq
with arbitrary sets of functions  $\phi _i(\mu )$. In the next section we
demonstrate that the quantity (\ref{taumiwa}) is exactly a KP $\tau $-function
in Miwa coordinates, and we return to the Kac-Schwarz operator in sect.5.

\subsection{KP $\tau $-function in Miwa parameterization}

A generic KP $\tau $-function is a correlator of a special form
\cite{DJKM}:

\beq
\tau ^G\{T_n\} = \langle 0|:e^{\sum \ T_nJ_n}: G|0\rangle
\eeq
with

\beq\label{ff}
J(z) = \tilde \psi (z)\psi (z)\hbox{; }  G =\
:\exp \ {\cal G}_{mn}\tilde \psi _m\psi _n:
\eeq
in the theory of free 2-dimensional fermionic fields  $\psi (z)$,
$\tilde \psi (z)$  with the action $\int
\tilde \psi \bar \partial \psi $. The vacuum states are defined by conditions

\beq
\psi _n|0\rangle  = 0\ \ n < 0\hbox{ , }  \tilde \psi _n|0\rangle  = 0\ \ n
\geq  0
\eeq
where  $\psi (z) =
\sum _{\bf Z}
\psi _nz^n\ dz^{1/2} $ , $\tilde \psi (z) =
\sum _{\bf Z}
\tilde \psi _nz^{-n-1}\ dz^{1/2}$.

The crucial restriction on the form of the correlator, implied by
(\ref{ff}) is
that the operator  $:e^{\sum \ T_nJ_n}:$  $G$  is {\it Gaussian} exponential,
so that the insertion of this operator may be considered just as a
modification of  $\langle \tilde \psi \psi \rangle$  {\it propagator}, and
the Wick theorem is applicable. Namely, the correlators

\beq\label{wick1}
\langle 0| \prod _i
\tilde \psi (\mu _i)\psi (\lambda _i) G|0\rangle
\eeq
for {\it any} relevant $G$ are expressed through the pair correlators of the
same form:

\beq\label{wick2}
(\ref{wick1}) =
{\rm det} _{(ij)} \langle 0| \tilde \psi (\mu _i) \psi (\lambda _j)
G|0\rangle
\eeq

The simplest way to understand what happens to the operator  $e^{\sum T_nJ_n}$
after the substitution of (\ref{miwa})
is to use the free-{\it boson}
representation
of the current  $J(z)=\partial \varphi (z)$. Then  $\sum  T_nJ_n =
\displaystyle {\sum _i
\left\lbrace  \sum _n {1\over n\cdot \mu _i^n}
\varphi _n\right\rbrace}  = \sum _i \varphi (\mu _i)$, and

\beq
:e^{\sum _i\varphi (\mu _i)}: = {1\over \prod _{i<j}(\mu _i-\mu _j)}
\prod _i :e^{\varphi (\mu _i)}:\hbox{ .}
\eeq
In fermionic representation it is better to start from

\beq
T_n = {1\over n} \sum _i ({1\over \mu ^n_i} - {1\over \tilde \mu _i^n} )
\eeq
instead of (\ref{miwa}). Then

\beq
:e^{\sum T_nJ_n}: = {\prod _{i,j}^N(\tilde \mu _i-\mu _j)\over
\prod _{i>j}(\mu _i-\mu _j) \prod _{i>j}(\tilde \mu _i-\tilde \mu _j)}
\prod _i \tilde \psi (\tilde \mu _i)\psi (\mu _i)\hbox{ .}
\eeq
In order to come back to (\ref{miwa}) it is necessary to shift all
$\tilde \mu _i$'s  to infinity. This may be expressed by saying that the left
vacuum is substituted by

$$
\langle N| \sim
\langle 0|\tilde \psi (\infty )\tilde \psi '(\infty )...\tilde \psi ^{(N-1)}(
\infty ).
$$
The $\tau $-function now can be represented in the form:

\beq
\new
\begin{array}{l}
\tau ^G_N[M] = \langle 0|:e^{\sum T_nJ_n}:G|0\rangle  =
\Delta (M)^{-1}\langle N|\prod _i
:e^{\varphi (\mu _i)}: G|0\rangle  = \\
=\lim _{\tilde \mu _j \to \infty}{\prod _{i,j}(\tilde \mu _i-\mu _j)
\over \prod _{i>j}
(\mu _i-\mu _j) \prod _{i>j}(\tilde \mu _i-\tilde \mu _j)}\langle 0|\prod _i
\tilde \psi (\tilde \mu _i)\psi (\mu _i)G|0\rangle
\end{array}
\eeq
applying the Wick's theorem (\ref{wick1}), (\ref{wick2}) and taking the
limit $\tilde \mu _i \to \infty$ we obtain:

\beq
\tau ^G_N[M] =
{{{\rm det} \ \phi _i(\mu _j)}\over {\Delta (M)}}
\eeq
with functions

\beq\label{2point}
\phi _i(\mu ) \sim  \langle 0|\tilde \psi ^{(i-1)}(\infty )\psi (\mu )
G|0\rangle
\stackreb{\mu \to
\infty}{\to} \mu ^{i-1}(1 + {\cal O}({1\over \mu})).
\eeq
Thus, we proved that KP $\tau $-function in Miwa coordinates (\ref{miwa}) has
exactly the determinant form (\ref{det}), or is a $\tau $-function of KP
hierarchy.

\subsection{Universal ${\cal L}_{-1}$-constraint and string equation}

Let us return to the question of specifying particular ''stringy"
solutions to the KP hierarchy which we already demostrated considering basis
vectors (\ref{ks}).  We will show that the matrix version of the Kac-Schwarz
operator which is almost

\beq\label{trlam}
Tr{\partial \over \partial \Lambda _{tr}} = Tr {1\over  W'(M)}
{\partial \over \partial M_{tr}}
\eeq
acting on $\tau$-function gives the string equation. Therefore it is natural to
examine, how this operator acts on

\beq\label{z}
Z^{\{ V\}}[M] = {{\rm det} \
\tilde \Phi _i(\mu _j)\over \Delta (M)}\prod _i
s(\mu _i),
\eeq

\beq
s(\mu ) = ( W'(\mu ))^{1/2}e^{ V(\mu )-\mu  W(\mu )},
\eeq

$$
\tilde \Phi _i(\mu ) = F_i(\lambda ) =
(\partial /\partial \lambda )^{i-1}F_1(\lambda )$$
First of all, if $Z^{\{ V\}}$ is considered as a function
of  $T$-variables,

\beq
{1\over Z^{\{ V\}}}Tr{\partial \over \partial \Lambda _{tr}}
Z^{\{ V\}} = -\sum _{n\geq 1}Tr [{1\over  W'(M)M^{n+1}}]
{\partial logZ^{\{ V\}}\over \partial T_n}.
\eeq
On the other hand, if we apply (\ref{trlam}) to explicit formula
(\ref{z}), we obtain:

\beq
\new
\begin{array}{c}
{1\over Z^{\{V\}}}Tr{\partial \over \partial \Lambda _{tr}}
Z^{\{ V\}}\\
 = - Tr\ M + {1\over 2} \sum _{i,j}{1\over
 W'(\mu _i) W'(\mu _j)}
{ W'(\mu _i)- W'(\mu _j)\over \mu _i - \mu _j} +
Tr{\partial \over \partial \Lambda _{tr}}\log \ {\rm det} \ F_i(\lambda _j),
\end{array}
\eeq
We can prove that

\beq\label{se}
{{1\over Z^{\{ V\}}}{\cal L}_{-1}Z^{\{ V\}}} = -
{\partial \over \partial T_1}\log \ Z^{\{ V\}} + TrM -
Tr{\partial \over \partial \Lambda _{tr}}\log \ {\rm det} \ F_i(\lambda _j).
\eeq
can be used in order to suggest the formula for the
universal operator  ${\cal L}_{-1}$.

Here

\beq\label{l-1}
{\cal L}_{-1} =\sum _{n\geq 1}Tr
[{1\over  W'(M)M^{n+1}}] {\partial \over \partial T_n} + \nn \\
+ {1\over 2}
\sum _{i,j}{1\over W'(\mu _i) W'(\mu _j)}{ W'
(\mu _i)-
 W'(\mu _j)\over \mu _i - \mu _j} - {\partial \over \partial T_1},
\eeq
So, in order to prove the ${\cal L}_{-1}$-constraint, one
should prove that the $r.h.s$. of (\ref{se}) vanishes, i.e.

\beq\label{dt1}
{\partial \over \partial T_1}\log \ Z^{\{ V\}}_N = TrM -
Tr{\partial \over \partial \Lambda _{tr}}\log \ {\rm det} \ F_i(\lambda _j)
\hbox{,}
\eeq
This is possible to prove only if we remember that
$Z^{\{ V\}}_N = \tau ^{\{ V\}}_N$. In this case the
$l.h.s$. may be represented as residue of the ratio

\beq\label{res}
res_\mu {\tau ^{\{ V\}}_N(T_n+
\mu ^{-n}/n)\over \tau ^{\{ V\}}_N(T_n)} =
{\partial \over \partial T_1}\log \ \tau ^{\{ V\}}_N(T_n).
\eeq
However, if expressed through Miwa coordinates, the $\tau $-function in the
numerator is given by the same formula with one {\it extra} parameter  $\mu $ ,
$i.e$. is in fact equal to $\tau ^{\{ V\}}_{N+1}$ . This idea is almost
enough to deduce (\ref{dt1}). For example, if
$N = 1$

$$
\tau ^{\{ V\}}_1(T_n) = \tau ^{\{ V\}}_1[\mu _1] =
e^{ V(\mu _1)-\mu _1 W(\mu _1)}[{ W'}(\mu _1)]^{1/2}F(
\lambda _1){ , }
$$

\beq\label{315}
\new
\begin{array}{c}
\tau ^{\{ V\}}_1(T_n+\mu ^{-n}/n) =
\tau ^{\{ V\}}_2[\mu _1,\mu ] = \\
= e^{ V(\mu _1)-\mu _1 W(\mu _1)}e^{ V(\mu )-\mu
 W(
\mu )} {[ W'(\mu _1) W'(\mu )]^{1/2}\over \mu  -
\mu _1}[F(\lambda _1)\partial F(\lambda )/\partial \lambda  -
F(\lambda )\partial F(\lambda _1)/\partial \lambda _1] = \\
= {e^{V(\mu )-\mu  W(\mu )}[ W'(\mu )]^{1/2}
F(\lambda )\over \mu  - \mu _1} \tau ^{\{ V\}}_1[\mu _1]\cdot [ -
\partial logF(\lambda _1)/\partial \lambda _1 +
\partial logF(\lambda )/\partial \lambda ].
\end{array}
\eeq
The function

\beq
F(\lambda ) =\int   dx\ e^{- V(x)+\lambda x} \sim
e^{ V(\mu )-\mu  W(\mu )} [ W'(\mu )]^{-1/2}\{1 +
O({ W'''\over  W' W'})\}.
\eeq
If  $ W(\mu )$  grows as  $\mu ^p$ when  $\mu  \rightarrow  \infty $ ,
then
$ W'''/( W')^2 \sim  \mu ^{-p-1}$ , and for our purposes it is
enough to have  $p>0$ , so that in the braces at the $r.h.s$. stands
$\{1+o(1/\mu )\}  (\mu \cdot o(\mu ) \rightarrow  0$ as $\mu
\rightarrow  \infty )$. Then numerator at the $r.h.s$. of (\ref{315}) is
$\sim  1 + o(1/\mu )$, while the second item in square brackets behaves as
$\partial logF(\lambda )/\partial \lambda  \sim  \mu (1+o(1/\mu ))$.
Combining all this, we obtain:

\beq
{\partial \over \partial T_1}\log \ \tau ^{\{ V\}}_1 =
res_\mu \left\lbrace  {1+o(1/\mu )\over \mu  - \mu _1} [-
\partial logF(\lambda _1)/\partial \lambda _1 +
\mu (1+o(1/\mu ))]\right\rbrace
 =  \mu _1 - \partial logF(\lambda _1)/\partial \lambda _1.
\eeq
$i.e.$ (\ref{dt1}) is proved for the particular case of  $N =1.$

In the particular case of monomial potential $V  =
{X^{p+1}\over p+1}$ (\ref{l-1}) turns into more common form
\cite{FKN1,DVV1}:

\beq\label{l-1p}
{\cal L}^{\{p\}}_{-1} ={1\over p}\sum _{n\geq 1}
(n+p)T_{n+p} {\partial \over \partial T_n} + \nn \\
+ {1\over 2p}
\sum _{a+b=p \atop{a,b \geq 0}}aT_abT_b
- {\partial \over \partial T_1},
\eeq

\section{Canonical quantization and p-q duality}

\subsection{General ideology}

Now, let us turn to somewhat more general question of  how a generic string
theory (first-
quantized or second quantized) should look like. In the simplest case
of topological string we can reduce ourselves to the question of basic
module space. In the frames of this ideology module spaces corresponding
to topological theories should be considered as a background for the
first-quantized theory while the second-quantized theory should be
related to the quantization of module space.

In the well-known case of pure topological gravity we should expect
{\it nothing}
since that theory does not have any target space at all. This is somehow
consistent with the observation that the partition function can be made
trivial just by a choice of gauge (polarization).

We are going to demonstrate that the matrix model solution can be obtained
within the frames of second quantization on a kind of ``module space" for these
theories (see \cite{Kri} for more detailed information on this point).

Finally we will make some comments on the considered problem in the framework
of mirror symmetry (see for example \cite{VafMir}). The important remark
is that
mirror manifolds should be distinguished classically and this effect is very
closely related to that one we have in the case of $ (p,q)$ models.

\subsection{String equation and Heisenberg algebra}

Now we are going directly to a problem of
description of a particular representation of the Heisenberg algebra.
One should start from \cite{Kri} where the ``phase space" for $(p,q)$
models is considered as a certain ``generalized" module space for the
Riemann surfaces with punctures. In the simplest case of sphere with
the only puncture one might take the phase space with a symplectic
structure

\beq\label{PB}
\{ W,Q\} = 1
\eeq
which is actually generated by

\beq
\{ z,t_1\} = 1
$$
$$
\{\tilde z, \tilde t_1\} = 1
\eeq
(where $z^p = W(\mu )$ and $\tilde z^q = Q(\mu )$).
For trivial $(1,p)$ topological theories $\tilde z = \mu $.

{}From this point of view what we consider is a quantization of a symplectic
manifold

\beq
\omega = \delta W \wedge \delta Q
 = \delta z \wedge \delta t_1
\eeq
and we can perform it by standard methods.

The corresponding action is

\beq
S = \int WdQ + S_0
$$
$$
dS = \delta W \wedge \delta Q
\eeq
and $S_0$ parameterizes an ``initial point". Now, it is obvious that in
the proposed quantization scheme the set of coupling constants depends
on the way of quantization, so does the solutions (potentials) of the
hierarchy, $\tau $- or the BA function {\it etc}.

Now the quantization gives the representation of the Heisenberg operators,
satisfying the string equation

\beq\label{ste}
[\hat P,\hat Q] = 1
\eeq
in the ``momentum" (spectral) space

\begin{equation}\label{repr}
\hat P =  \lambda
$$
$$
\hat Q = {\partial \over \partial \lambda } + Q(\lambda )
\end{equation}
{}From the point of view of the KP hierarchy, we will also add some additional
requirements on the ``spectral parameter" implying that

\begin{equation}
\lambda  = W(\mu ) = \mu ^p
\end{equation}
then $(p,q)$ models correspond to the case where  $Q(\lambda )$  should be a
{\it polynomial} of $\mu $ of degree $q$ \cite{FKN1}, (while
the corresponding wave functions should have specific asymptotics when  $\mu
\rightarrow  \infty $).

Wave functions  of this problem appear to be the Baker-Akhiezer functions
of the corresponding integrable system and
when acting on wave functions conditions (\ref{repr}) get the form of the
Kac-Schwarz equations \cite{KS,S}:

\begin{equation}\label{KS}
\lambda \varphi _i(\mu ) = \sum  _j W_{ij}\varphi _j(\mu )
$$
$$
\hat A\varphi _i(\mu ) = \sum  _j A_{ij}\varphi _j(\mu )
\end{equation}
where

\begin{equation}
\lambda  = W(\mu ) \sim  \mu ^p
$$
$$
A^{(W,Q)} \equiv  s^{(W,Q)}(\mu ){1\over W'(\mu )}
{\partial \over \partial \mu } [s^{(W,Q)}(\mu )]^{-1} =
$$
$$
= {1\over W'(\mu )} {\partial \over \partial \mu } - {1\over 2}
{W''(\mu )\over W'(\mu )^2} + Q(\mu )
\end{equation}
The standard way to construct wave functions of the theory is to
define the Fock vacuum by

\begin{equation}\label{Fock}
\hat A\Psi _0= 0
\end{equation}
with an obvious solution

\begin{equation}
\Psi _0 = \sqrt{W'(\mu)}\exp{\int{QdW}}
\end{equation}
and the corresponding $\tau $-function is a determinant
projection of  higher states

\begin{equation}
\Psi _n \sim W^n \Psi _0
\end{equation}
to the states with a {\it canonical} asymptotics

\begin{equation}
\varphi _i(\mu )
\stackreb{\mu \rightarrow \infty }{\rightarrow} \mu ^{i-1}
\end{equation}
forming the conventional basis in the space of wave functions -- the point of
infinite-dimensional Grassmannian.

The only simple case arises when the Kac-Schwarz equations
(\ref{KS}) have trivial solution, {\it i.e.} when  $p=1$.
Starting from normalization
$\varphi _1(\mu ) = 1$ (corresponding to $\Psi_0 = \exp{\int{Qd\mu}}$), and
using first of eqs.(\ref{KS}) one can always get
$\Psi_n = \mu^n\exp{\int{Qd\mu}} \rightarrow \varphi _i(\mu ) = \mu ^{i-1}$
{\it exactly}. Then the second condition of
(\ref{KS}) is fulfilled {\it automatically} for {\it any} $Q(\mu )$.

However, one can see that the corresponding solutions are related to
topological models by a kind of Fourier
transformation. Indeed, it has been observed
\cite{KM1,KM2} that the system of equations (\ref{KS}) posseses a {\it duality}
symmetry which relates $(p,q)$ to $(q,p)$ solution.
The duality transformation for the Baker-Akhiezer functions looks like

\begin{equation}
\psi ^{(P,Q)}(z) = [P'(z)]^{1/2}\int   dQ\ e^{P(z)Q(x)}\psi ^{(Q,P)}(x)
[Q'(x)]^{-1/2}
\end{equation}
and it can be also written for the basis vectors in the Grassmannian

\beq\label{dual}
\phi _i(\mu ) = [W'(\mu )]^{1/2} \exp ( - \left.S_{W,Q}\right |
_{x=\mu })  \int   d{\cal M}_Q(x)f_i(x) \exp \ S_{W,Q}(x,\mu )
\eeq
with

\begin{equation}\label{act}
d{\cal M}_Q(x) = dx \sqrt{Q'(x)}
$$
$$
S_{W,Q}(x, \mu) = - \int_{}^{x}{WdQ} + W(\mu)Q(x)
\end{equation}
and for the partition functions

\beq\label{dualpf}
\tau^{(W,Q)}\,\,[M] =
$$
$$
= C[V,M]\int DX\tau^{(Q,W)}\,\,[X]\exp \left \{Tr[1/2 \log Q'(X) +
\int_{M}^{X}W(z)dQ(z) + W(M)Q(X)] \right \}
\eeq
(here, better to consider {\it normalized} partition function
$\tau ^{(W,Q)} \rightarrow Z^{(W,Q)} \rightarrow
\Psi _{BA}^{(W,Q)}(t_k - {1 \over k}Tr M^{-k}$).
It makes possible to obtain solutions for
nontrivial models -- topological $(p,1)$
models \cite{KMMMZ} and their Landau-Ginzburg deformations \cite{LGGKM}.

\begin{equation}\label{LanGin}
\varphi _i(\mu ) = \sqrt{p\mu ^{p-1}}\exp \left(- \sum t_k\mu ^k\right)\int
dx\ x^{i-1}
\exp (-V(x) + x\mu ^p)
\end{equation}
which are {\it dual}  to $(1,p)$ model in the above sense.

Here, we immediately run into a puzzle: how to interpret this from the point
of view of quantization theory. Indeed, the duality transformation (\ref{dual})
is nothing but a transformation from $\hat p$ to $\hat q$ quantization
procedure or from one to another representation of quantum algebra and as it
is well-known the quantization should be independent of this. It means that
$(p,1)$ and $(1,p)$ or trivial theory are in fact equivalent as string
theories, i.e. the nontrivial partition functions for $(p,1)$ theories
corresponding to some well-known topological theories (twisted
$N=2$ Landau-Ginzburg theories) give nothing from a physical point of view
\footnote{$(2,1)$
model corresponds to pure topological gravity and generates intersection
indices on module spaces of Riemann surfaces with punctures - it appears that
the intersection indices in topological gravity are just a "physical
artefact"}.
Thus, the first puzzle is that $\tau ^{(1,p)} \equiv 1$ seems to contain all
the ``topologcal" information as a ``dual" partition function does. Second, the
topological numbers perhaps should not be considered as "real observables" of
the theory -- they rather correspond to a sort of combinatorial factors for
Feynman diagramms in particle theories.

This is actually a new feature of string theory if we compare it to quantum
field theory -- i.e. even trivial target-space model can possess rich and
nontrivial structure. The Virasoro action in these theories naturally follows
from (\ref{repr}), (\ref{KS}).

Let us finally add few comments about holomorphic anomaly. The
``quasiclassical" $\tau $-function obeys a homogeneous relation

\beq
\sum t_j {\partial \over \partial t_j } \log \tau _0 = 2 \log \tau _0
\eeq
spoilt by the contribution of the one-loop correction, having the form,
for example, for the $(2,1)$ theory

\beq\label{holan}
\sum t_j {\partial \over \partial t_j } \log \tau  - 2 \log \tau =
- {1\over 24}
\eeq
The similiar expressions appear when one considers the logariphm of the
partition function for the higher-dimensional theories \cite{holan} and this
should mean that the expression (\ref{holan}) should have a similiar nature.

\section{Conclusion}

Now let us briefly summarize the main ideas presented above. We have tried to
demonstrate that appearing in the context of matrix models effective
target-space description of string theory can be a useful tool for constructing
a nonperturbative string field theory. Indeed, the space of coupling constants
$ \{ T_k \}$ may be considered for simplest string models as a space of
background fields and one might hope to get a second-quantized theory by
quantization of appearing there structures. It has been shown by Krichever
\cite{Kri} that the "small phase space" in fact can be considered as a
certain module
space for a spectral surface with marked points if one restricts the order of
singularities in these marked points. Then it is natural to consider the
quantization of (\ref{PB}) as a quantization of this module space. In fact we
have shown above that the particular example of $ (p,1)$ models rather leads to
a trivial theory -- topological gravity (W-gravity) which is not too much
interesting as a target space theory. However, the natural question that
appears is a generalization of this approach to more interesting module spaces.

For example, there exists a quite interesting scheme of quantization of module
spaces of flat connections and projective structures on Riemann surfaces
with punctures \cite{FR}. This is not far from what
we need in the case of string models: in fact module spaces of flat connections
already appeared in the context of two-dimensional Yang-Mills theory and its
relation to string theory \cite{GT,GoNe}. It is natural to think that the
related string models should have partition (generating) functions more simple
than the discussed above theories, being related thus from the point of view of
integrable hierarchies with the, say, rational $ \tau$-functions. The
appearence of such $ \tau$-functions can be interpreted
in the way that a restricted amount of world-sheet topologies give contribution
to the partition function. In fact \cite{GKM2} there exists another, so-called
"character" phase of GKM considered above which is closely related to the
Yang-Mills theory and rational $\tau$-functions.

\section{Acknowledgements}

I would like to thank V.Fock, A.Gerasimov,  A.Gorsky, R.Kashaev, S.Kharchev,
I.Krichever, A.Levin, A.Losev, A.Mironov, A.Morozov, A.Niemi, K.Palo, A.Rosly
and V.Rubtsov for fruitful discussions.

The work was in part supported by Fundamental Research Foundation of Russia,
contract No 93--02--3379 and by NFR-grant No F-GF 06821-305 of the Swedish
Natural Science Research Council. I am grateful for warm hospitality to the
organizers of the Third Baltic Rim student seminar and to the Institute of
Theoretical Physics of Uppsala University when this paper has been completed.

\end{document}